\documentclass[osajnl,twocolumn,showpacs,superscriptaddress,11pt]{revtex4-1} 

\usepackage{epsfig} 
\usepackage{color} 
\usepackage{mathrsfs}
\usepackage{amsmath}
\usepackage{amssymb}
\usepackage{amsfonts}
\usepackage{mathrsfs}
\usepackage{color}
\usepackage{tikz}
\usepackage[pdfa]{hyperref}

\pgfdeclarelayer{foreground}
\pgfdeclarelayer{background}
\pgfsetlayers{background,main,foreground}

\begin{document}

\newcommand{\mnras}{ Mon. Not. R. Astron. Soc.}
\newcommand{\amp}{\&}
\newcommand{\pasp}{Publ. Astron. Soc. Pac.}
\newcommand{\aeta}{Astron. Astrophys.}
\newcommand{\aap}{\aeta} 
\newcommand{\aaps}{Astronomy and Astrophysics Supplement}
\newcommand{\optcom}{Optics Communications}
\newcommand{\aplopt}{Applied Optics}
\newcommand{\optlet}{Optics Letters}
\newcommand{\solphys}{Solar Physics}

\newcommand{\real}{\mathrm Re}
\newcommand{\ft}[1]{\mathscr{F}\left\{#1\right\}}
\newcommand{\npd}{M}
\newcommand{\nwfs}{M'}
\newcommand{\conv}{\ast}
\newcommand{\half}{0.5}
\newcommand{\quar}{0.25}
\newcommand{\affi}{A_t}

\title{Calibrating a high-resolution wavefront corrector with a
  static focal-plane camera}


\author{Visa Korkiakoski} 
\affiliation{Leiden Observatory, Leiden University, P.O. Box 9513, 2300 RA Leiden, The Netherlands}
\email{korkiakoski@strw.leidenuniv.nl}

\author{Niek Doelman}
\affiliation{TNO Technical Sciences, Stieltjesweg 1, 2628 CK Delft, The Netherlands}
\affiliation{Leiden Observatory, Leiden University, P.O. Box 9513, 2300 RA Leiden, The Netherlands}

\author{Johanan Codona}
\affiliation{Steward Observatory, University of Arizona, Tucson, AZ 85721, USA}

\author{Matthew Kenworthy}
\affiliation{Leiden Observatory, Leiden University, P.O. Box 9513, 2300 RA Leiden, The Netherlands}

\author{Gilles Otten}
\affiliation{Leiden Observatory, Leiden University, P.O. Box 9513, 2300 RA Leiden, The Netherlands}

\author{Christoph U. Keller}
\affiliation{Leiden Observatory, Leiden University, P.O. Box 9513, 2300 RA Leiden, The Netherlands}


  
\begin{abstract}
  We present a method to calibrate a high-resolution
  wavefront-correcting device with a single, static camera, located in
  the focal-plane; no moving of any component is needed. The method is
  based on a localized diversity and differential optical transfer
  functions (dOTF) to compute both the phase and amplitude in the
  pupil plane located upstream of the last imaging optics.  An
  experiment with a spatial light modulator shows that the calibration
  is sufficient to robustly operate a focal-plane wavefront sensing
  algorithm controlling a wavefront corrector with $\sim$40~000
  degrees of freedom. We estimate that the locations of identical
  wavefront corrector elements are determined with a spatial
  resolution of 0.3\% compared to the pupil diameter.
\end{abstract}







\maketitle

\section{Introduction}


In certain situations, such as exoplanet imaging, it is necessary to
have an extremely good wavefront (WF) quality.  To achieve this, the
wavefront must be corrected with a very high resolution; the
next-generation extremely large telescopes having apertures larger
than 20~m require deformable mirrors with up to $200\times 200$
actuators. Furthermore, additional WF sensing must be done at the
focal plane to avoid the slowly evolving, non-common path aberration
errors.

A possible WF reconstruction algorithm for this purpose is the Fast \&
Furious (F\&F) algorithm
\cite{keller2012spie,korkiakoski2012spie1}. It is numerically
extremely efficient, relying on small WF aberrations, pupil symmetries
and phase-diversity to achieve very fast WF reconstruction.

To validate the F\&F algorithm for high-order wavefront correction, we
use an inexpensive spatial light modulator (SLM), based on
twisted-nematic liquid crystals.  It has about $300\times 300$
wavefront-modifying pixels surrounding the aperture.  The device is
able to make a stroke of $\sim$1~rad while maintaining a sufficiently
uniform transmittance.

A successful operation of the system requires precise knowledge of how
the SLM reacts to the control signal. In addition, it is necessary to
know, with sub-pixel accuracy, where the SLM elements are located with
respect to the physical pupil. Information about the pupil amplitudes
is important for high-contrast applications, particularly when an SLM
causes amplitude aberrations.

Therefore, it is highly desirable to have a method that detects both
the phase and amplitude changes in the pupil plane without the need to
physically move the imaging camera or wavefront correcting elements.

We found that a single method, simple and easy to implement, is
sufficient for all our calibration purposes:  the differential
optical transfer function method (dOTF) \cite{codona2012,
  codona2012b, codona2013}.

In this paper, we show results from optical experiments demonstrating
how the dOTF method can be used for high-contrast imaging
calibrations. In Section~\ref{sec:theory} we describe the theoretical
background for the calibration method. Section~\ref{sec:methods}
discusses practical issues in the experiments, and
Section~\ref{sec:results} shows the results of the dOTF algorithm and
the high-resolution F\&F performance. Finally,
Section~\ref{sec:conclusions} draws the conclusions.

\section{Theoretical background}
\label{sec:theory}

The dOTF method is a phase-diversity technique that reconstructs the
electric field at the pupil-plane using intensity measurements at the
focal plane. However, the method is unique in not requiring any models
or a priori information: nothing has to be known about the pupil
function, wavefront corrector or the PSF sampling. On the contrary,
the dOTF method can be used to calculate, with good accuracy, the
parameters required by more conventional phase-diversity algorithms,
discussed, for instance, by \cite{gerchberg1972,fienup82}.

The key is a very localized diversity at the edge of the
pupil. Once the diversity is close to a delta function, it becomes
possible to directly extract the pupil-plane complex amplitudes from
the intensity measurements. However, this approach has a major
challenge: the image with a delta-function diversity is almost
identical to the original image.

The comparison of two very similar images results in issues with the
signal-to-noise ratio (SNR). To counteract the SNR problem, we found,
as discussed in Section~\ref{sec:noisecomp}, that it is extremely
helpful to apply a large defocus when applying the dOTF
method. Therefore, the following discussion is geared towards applying
the dOTF method for calibration purposes, where only the change of the
pupil-plane complex amplitudes is of interest and not the image itself.

We begin by describing the basic principles of the dOTF method in
Section~\ref{sec:basics}. Section \ref{sec:esterr} discusses the dOTF
accuracy in our setup, and Section~\ref{sec:sampling} explains how
to determine the PSF sampling with the dOTF method.

\subsection{Basic principle of dOTF method}
\label{sec:basics}

\begin{figure}[hbtp]  \center
\includegraphics[width=2.75cm]{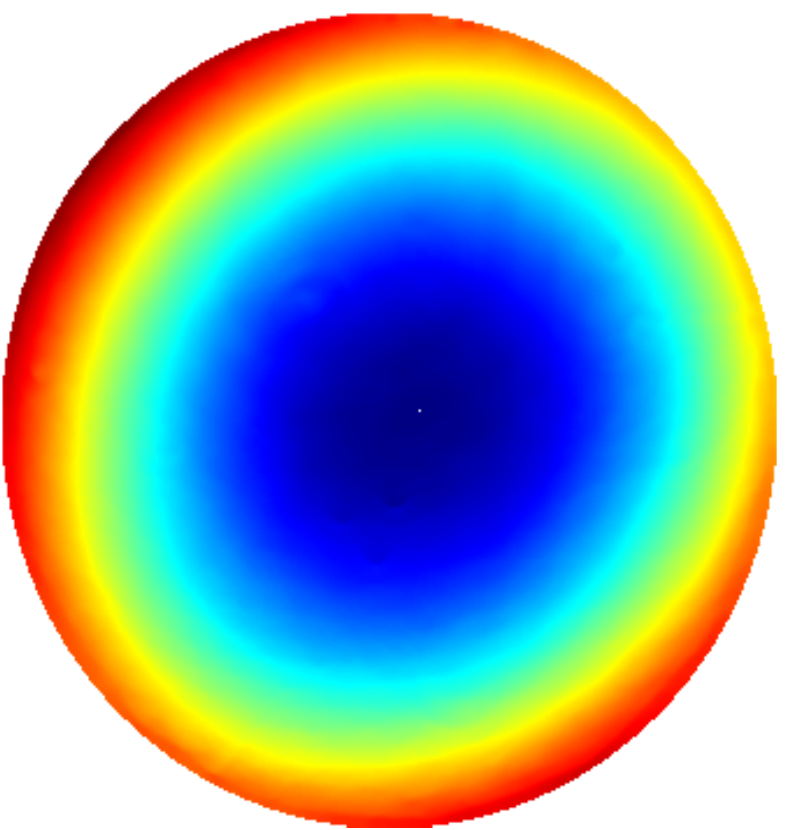}
\includegraphics[width=2.75cm]{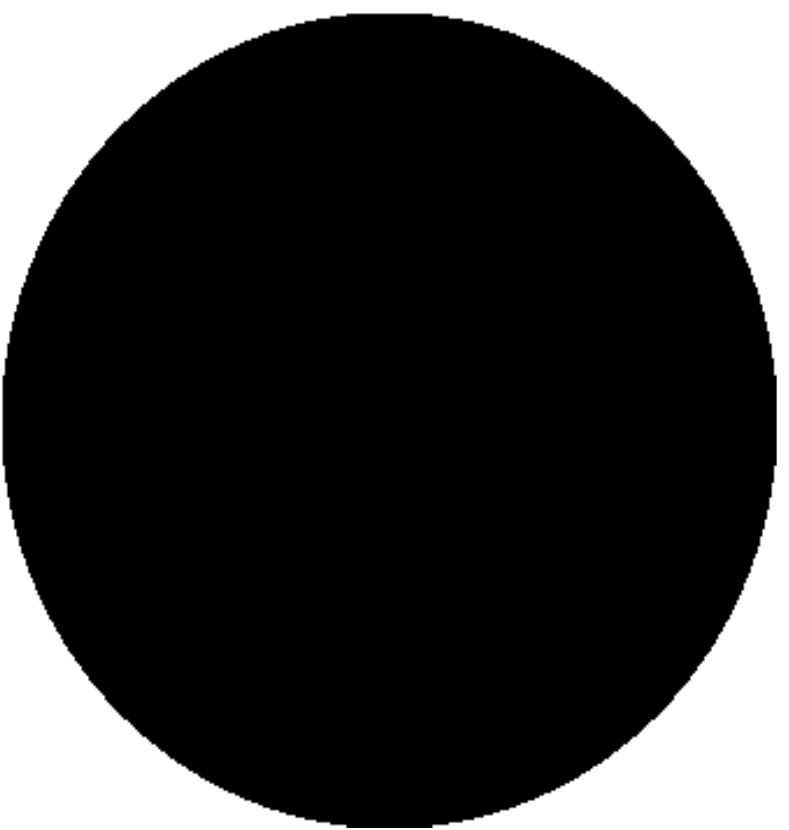}
\includegraphics[width=2.75cm]{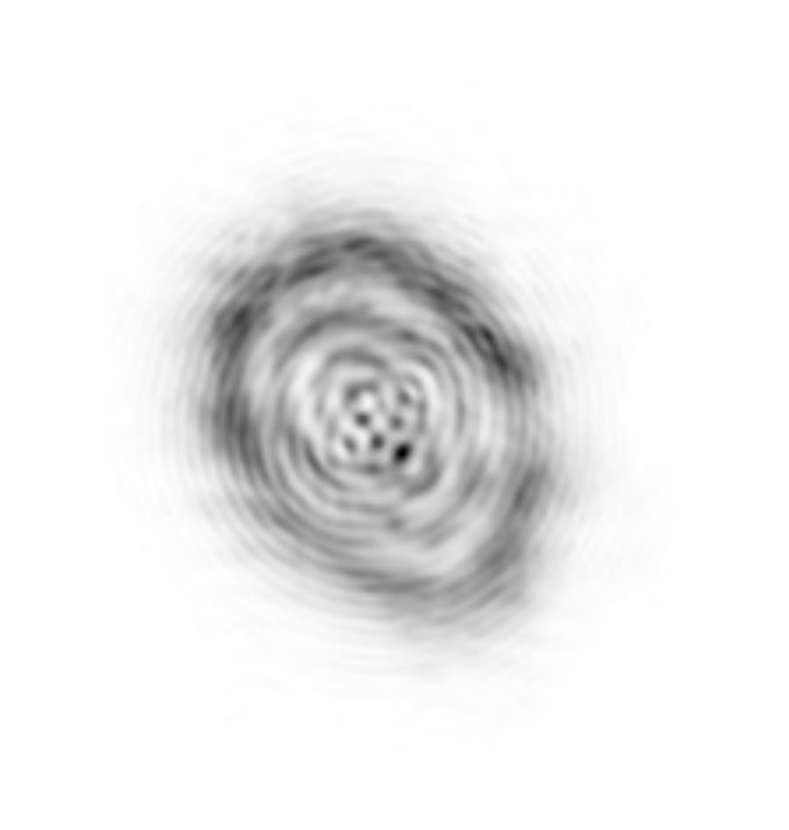}
\caption{A simulated example of a PSF. Left: wavefront having a large
  defocus, 30~rad peak-to-valley, on top of a high-order component
  with peak-to-valley $\sim$2~rad. The wavefront is obtained from
  laboratory measurements (see Section~\ref{sec:dotf0}), scaled to a
  size of $334\times 334$~pixels. Middle: perfect circular pupil
  having the same size. Right: the resulting image (shown area is
  $220\times 220$~pixels, the total simulated size is $1146\times
  1146$~pixels).}
\label{fg:calisample1}
\end{figure}

When imaging a monochromatic point source, the image can be modeled
using Fraunhofer diffraction. The image in the focal plane is the
squared modulus of the complex amplitudes in the pupil plane,
\begin{equation} \label{eq:p}
p = \ft{A \exp(i \phi)} \left[\ft{A \exp(i \phi)} \right]^*,
\end{equation}
where $\phi$ is the wavefront in the pupil, $A$ is the
absolute value of the electric field in the aperture (also describing 
the pupil shape), and $^*$ denotes the complex conjugate. An example 
of such an image is shown in Fig.~\ref{fg:calisample1}.

The optical transfer function is the Fourier transform of
Eq.~\eqref{eq:p} and can be written as
\begin{equation}
P = A \exp(i \phi) \conv A' \exp(-i \phi'),
\end{equation}
where $A'$ and $\phi'$ denote the mirrored versions of $A$ and $\phi$:
$A'(x,y)=A(-x,-y)$, $\phi'(x,y)=\phi(-x,-y)$. It is convenient to
consider the OTF as a convolution of complex functions,
\begin{equation}
P = \Psi \conv \Psi',
\end{equation}
where $\Psi$ denotes the complex amplitudes at the pupil plane, and
$\Psi'$ is its mirrored and conjugated variant. 

To solve the complex amplitudes using only intensity measurements, it
is necessary to introduce a diversity -- a modification in the
wavefront and/or pupil transmittance -- and to record another
image. The OTF of the diversity image can be written as
\begin{equation}
P_2 = (\Psi + \Psi_d) \conv (\Psi' + \Psi'_d),
\end{equation}
where $\Psi_d$ is the change in complex amplitudes in the pupil.

The difference of the the optical transfer functions is
\begin{equation}  \label{eq:pd1}
P_d = P_2 - P = \Psi \conv \Psi'_d + \Psi' \conv \Psi_d + \Psi_d \conv \Psi'_d.
\end{equation}

\begin{figure}[hbtp]  \center
\raisebox{-0.3cm}[2.8cm][0cm]{
\begin{tikzpicture}
\begin{pgfonlayer}{foreground}
     \path  (-2.9,-0.5) node (c) {{$N_1$}}
        (-0.6,0) node (b) {{$N_2$}};
     \draw [|-|] (-0.3,-1.85) -- (-0.3,1.85);
     \draw [|-|] (-2.5,0) -- (-2.5,-1.13);
\end{pgfonlayer}
\begin{pgfonlayer}{background}
 \path      (0,0) node (o) {
   \fbox{\includegraphics[width=3.7cm]{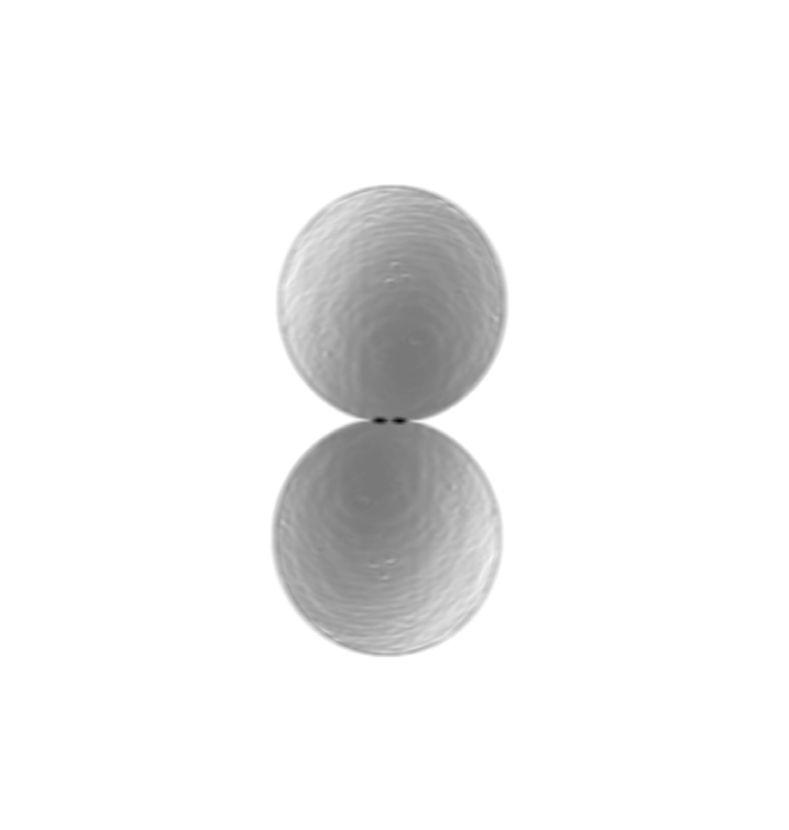}}
   \includegraphics[width=3.7cm]{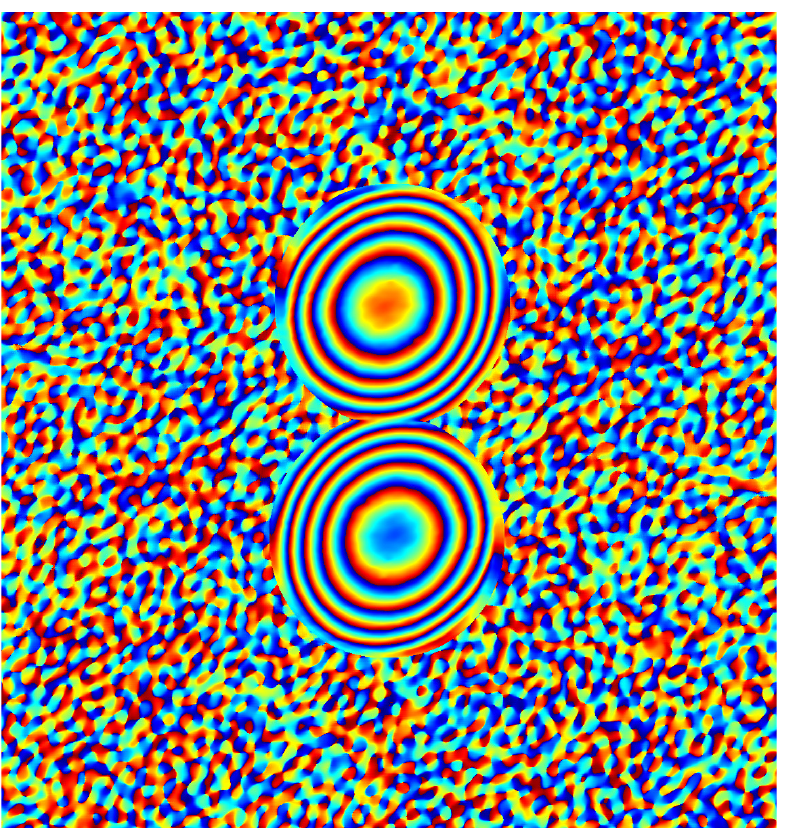}};
\end{pgfonlayer}
\end{tikzpicture}
}
\caption{A simulated dOTF ($P_d$). The array size is $1146\times
  1146$~pixels. Left: modulus -- $N_1$ indicates the pupil diameter in
  pixels, and $N_2$ is the size of the whole dOTF array. Right:
  phase.}
\label{fg:calisample3}
\end{figure}

If the diversity is highly localized, then $\Psi_d$ is approximately a
delta function (multiplied by a complex constant).  A simulated
example of such a difference is shown in Fig.~\ref{fg:calisample3}.
The wavefront and pupil are the same as in
Fig.~\ref{fg:calisample1}. The pupil width is 334 pixels, and the
diversity is a disk, 15 pixels wide, at the edge of the pupil.

\begin{figure}[hbtp]  \center
\fbox{\includegraphics[width=2.45cm]{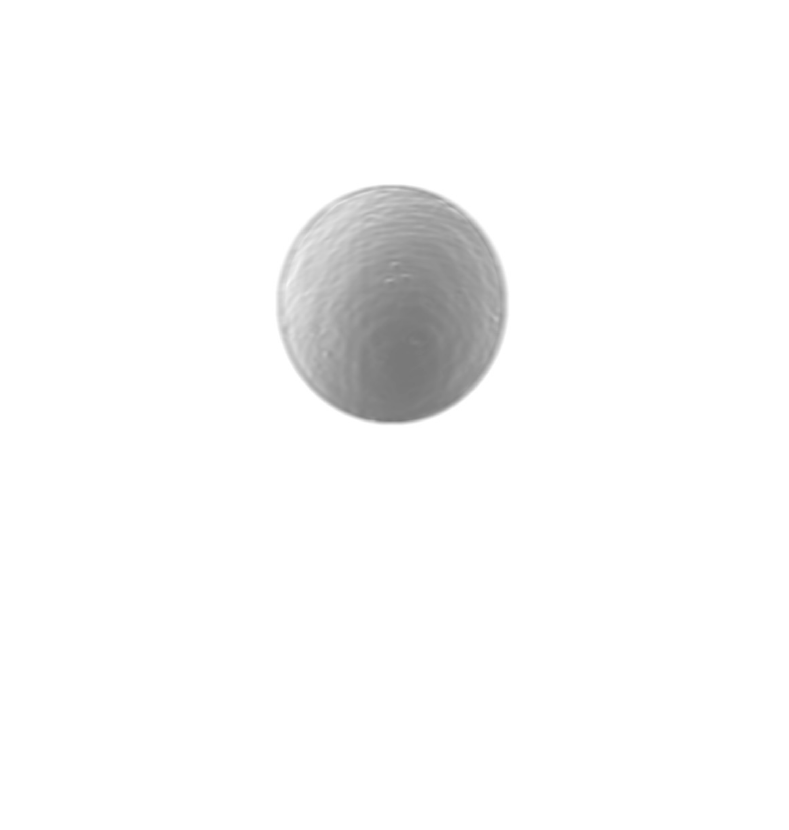}}
\fbox{\includegraphics[width=2.45cm]{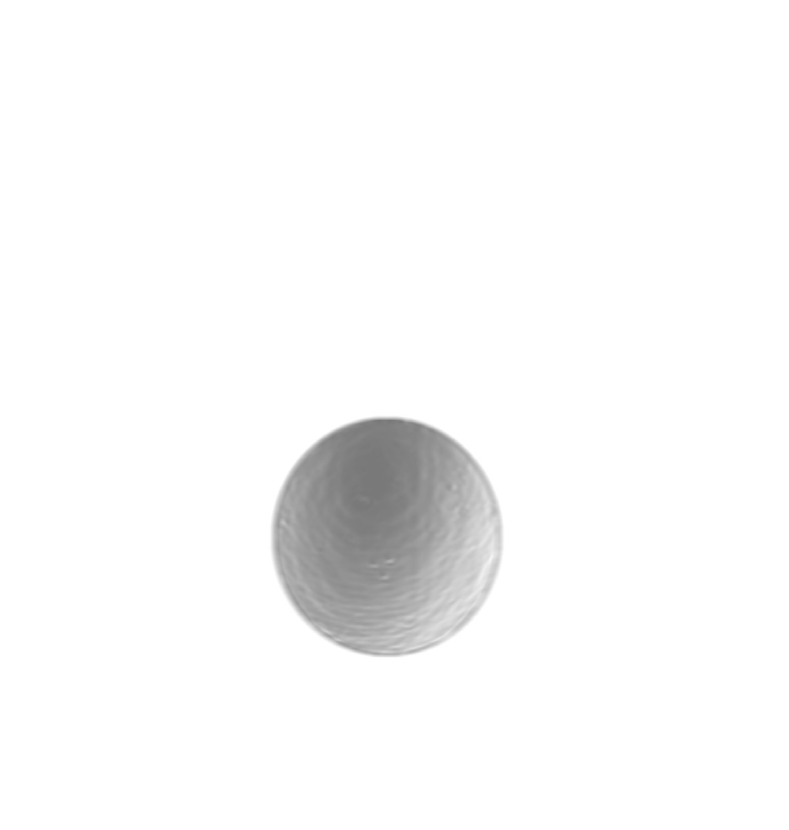}}
\fbox{\includegraphics[width=2.45cm]{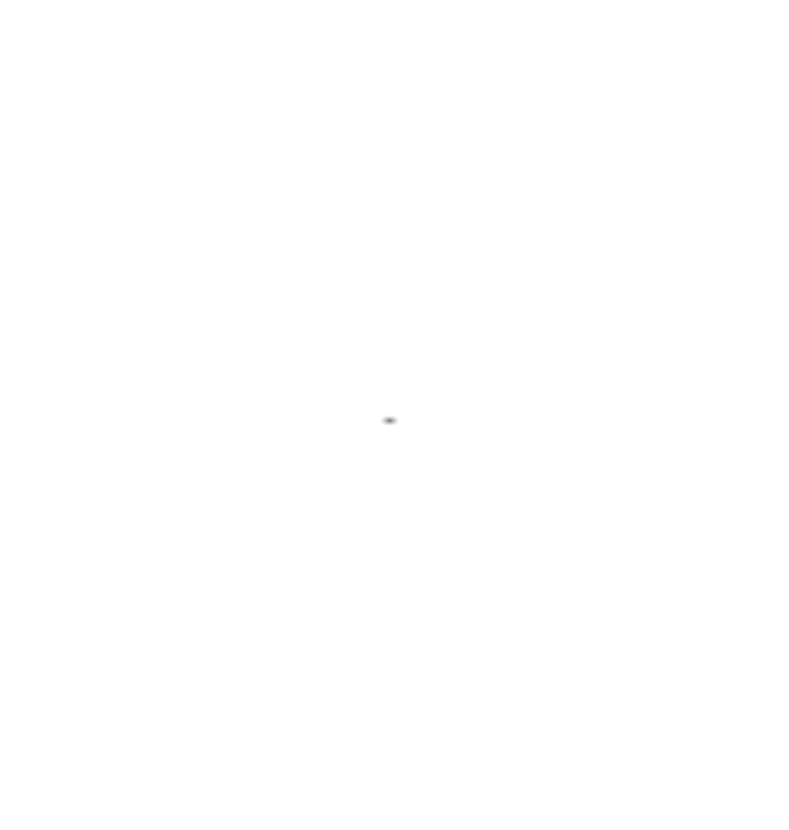}}
\caption{Moduli of the simulated terms in Eq.~\eqref{eq:pd1}: $\Psi
  \conv \Psi'_d$, $\Psi' \conv \Psi_d$ and $\Psi_d \conv \Psi'_d$.}
\label{fg:calisample2}
\end{figure}

Highly localized means that the range of the diversity is
less than $\sim$1--5\% of the pupil width; in such cases it has
nonzero values only close to itself. If the diversity is at the edge
of the pupil, the two first terms in Eq.~\eqref{eq:pd1} have nonzero
values at different locations -- except for close to the diversity.
This can be further illustrated by inspecting the different terms in
Eq.~\eqref{eq:pd1} separately, as is done in Fig.~\ref{fg:calisample2}.

The two pupil functions are shifted because the peaks of the
convolving diversity functions, $\Psi_d$ and $\Psi'_d$, are separated
by about one pupil diameter. The pupil amplitudes are somewhat
distorted, since $\Psi_d$ is not a perfect delta function.  The impact
of the third term, $\Psi_d \conv \Psi'_d$, is negligible in most
cases.

Thus, it is possible to consider Eq.~\eqref{eq:pd1} piecewise,
\begin{align}
P_{d}(x,y) &= \Psi \conv \Psi'_d  \approx  \Psi \conv \delta'_d,  &(x,y) \in D\\
P_{d}(x,y) &= \Psi' \conv \Psi_d  \approx  \Psi' \conv \delta_d,  &(x,y) \in D',
\end{align}
where $D$ and $D'$ denote the regions on opposite sides of the
diversity location, and $\delta_d$ and $\delta'_d$ are delta functions
multiplied by a complex constant.

\subsection{Analysis of dOTF based calibration}
\label{sec:esterr}

The measurement error of the dOTF method is almost entirely caused by
the smoothing convolution with $\Psi_d$ and noise propagation.
For our purposes, it is also necessary to have a more detailed look at
the error made when characterizing the performance of the SLM. We used
the SLM itself to generate the localized diversity, but a similar
effect could also be achieved by using an additional device
\cite{codona2013}.

A numerical example of the convolution bias in a typical measurement
situation was set up with a small, constant, WF difference of
0.5~rad and a transmittance reduction of 10\% introduced in one half
of the pupil; the undistorted case is the same as shown in
Fig.~\ref{fg:calisample1}. Two dOTF measurements (in total
four images) are recorded: both the undistorted and the modified
complex amplitudes are determined.  These two dOTF arrays are then
used to determine the change in the pupil plane. We measure the phase
change (denoted as $\Delta_\phi$) as a subtraction of reconstructed
wavefronts. The change in transmittance ($\Delta_A$) is measured as a
ratio of dOTF moduli. It holds that
\begin{align}
  \Delta_\phi &= \text{phase}\left({\Psi_2 \conv \Psi'_{d2}}\right) -
  \text{phase}\left({\Psi \conv \Psi'_{d}}\right) \\
  \Delta_A &= \frac{\left|\Psi_2 \conv \Psi'_{d2}\right|}{\left|\Psi
      \conv \Psi'_d\right|},
\end{align}
where $\text{phase}(\cdot)$ denotes a function unwrapping the phase of
a complex number, $\Psi_2$ is the true complex amplitude after
diversity modification, and $\Psi'_{d2}$ describes the localized
diversity of the dOTF measurement made for the modified case.  In our
tests, we always had high SNR, and therefore we did not observe 
issues with $\left|\Psi \conv \Psi'_d\right|$ being zero.

\begin{figure}[hbtp]  \center
\includegraphics[width=3.95cm]{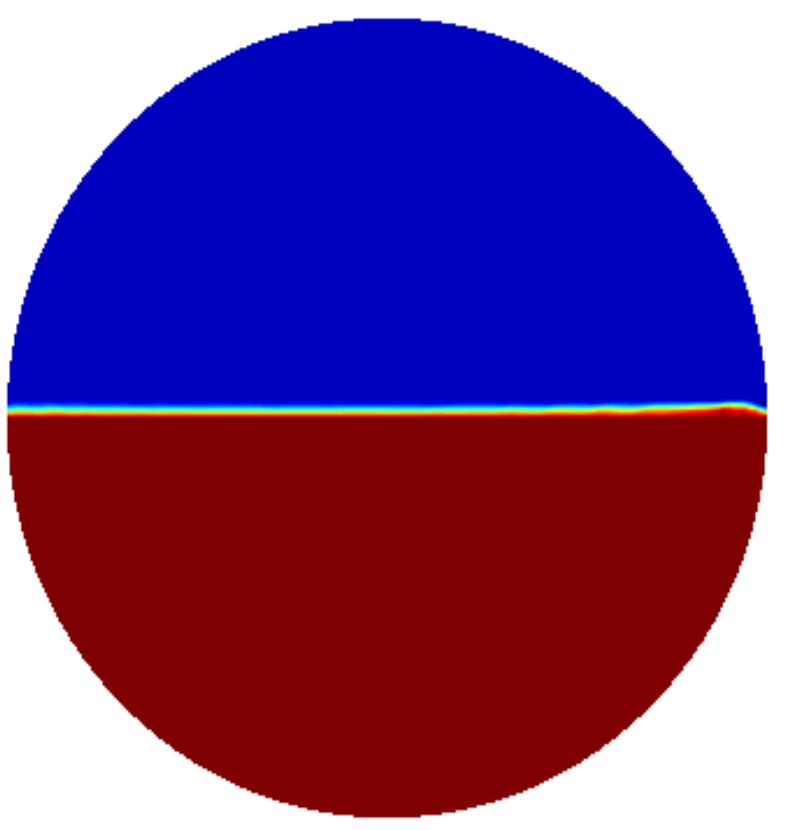}
\includegraphics[width=3.95cm]{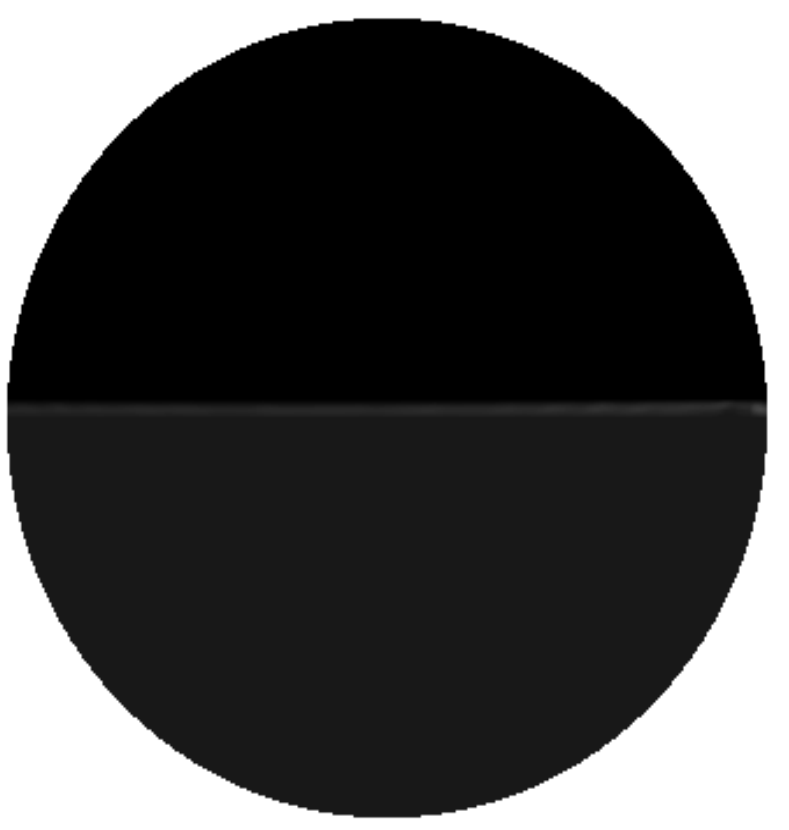}

\includegraphics{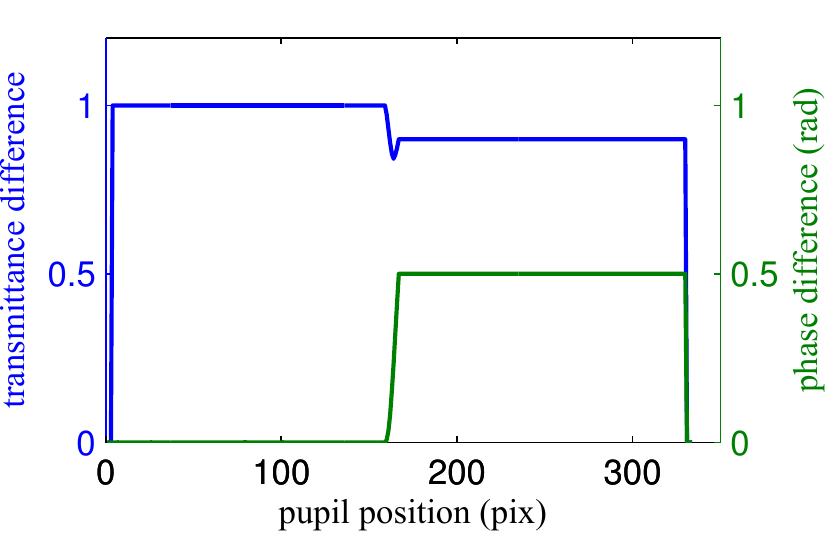}
\caption{Illustration of the change in the pupil plane, measured by the
  dOTF method. Top left: $\Delta_\phi$. Top right: $\Delta_A$. Bottom:
  vertical cuts of $\Delta_\phi$ and $\Delta_A$.}
\label{fg:errsimu}
\end{figure}

The arrays $\Delta_\phi$ and $\Delta_A$ are shown in
Fig.~\ref{fg:errsimu}.

Both the phase and transmittance changes are observed without any
bias. Only the convolution smoothing distorts the determined complex
amplitudes -- it is especially visible along the edge of the
modification. The edge appears as a linear transition in the phase,
but in the transmittance, the edge also has an undershoot of
$\sim$5\%.

It is particularly interesting that the transmittance change,
$\Delta_A$, shows no signs of the dOTF modulus distortions, visible
for instance in Figs.~\ref{fg:calisample3} and
\ref{fg:calisample2}. This obviously holds only when the introduced
phase and transmittance changes are small enough, but this was the
case with all the measurements presented in this paper -- the introduced
phase change never exceeded 1.6~rad.

The numerical example shown is not an extensive analysis of the dOTF
method; a more thorough approach would be needed to characterize
its limits, which is outside the scope of this paper.

\subsection{Sampling}
\label{sec:sampling}

Another important issue is the sampling at the detector array.
Typically, the sampling is expressed in terms of the width of the
diffraction-limited PSF, in pixels, and its value is determined both by
the optical design and the detector properties.

Determining the sampling accurately is important for all
model-dependent focal plane wavefront sensing techniques.  When the
PSF is modeled numerically with an FFT, its diffraction-limited width
is usually defined as
\begin{equation}  \label{eq:sample}
  w = \frac{N_{\text{arr}}}{N_{\text{pup}}},
\end{equation}
where $N_{\text{arr}}$ is the width of the zero-padded FFT array in
pixels, and $N_{\text{pup}}$ is the diameter of the modeled pupil in
pixels.  It can be difficult to determine values for $w$ that
accurately match the modeled system, if only distorted PSF images are
available.

However, the dOTF method offers a straightforward way to determine $w$
with an accuracy close to the dOTF array discretization limit; a greater
precision would require dealing with the fixed sampling of the FFT
\cite{zielinski2011phd}. First, the dOTF array is obtained by Fourier
transforming pre-processed and subtracted detector intensities. Then,
the value of $w$ is obtained directly: it is the ratio between the
total width of the dOTF array and the observed pupil diameter in
pixels -- indicated by $N_2$ and $N_1$ in Fig.~\ref{fg:calisample3}.

Obviously, a successful use of the dOTF method requires that the
detector array is at least Nyquist sampled (two pixels per
diffraction-limited PSF core) -- a limitation when compared to general
phase-retrieval methods based on error metric minimization
\cite{brady2009}. If the sampling was smaller ($w<2$), aliasing would
occur. Its concrete manifestation would be that the joined pupils (as
shown in Fig.~\ref{fg:calisample3}) would not fit into the FFT array,
and this would render the dOTF method useless. To avoid this with a
safe margin, we used a moderate oversampling, $w\approx 3$. A larger
over-sampling would work as well, but it would require more detector
pixels and bigger arrays to handle the data.


\section{Experimentals considerations}
\label{sec:methods}

We explain the experimental arrangements we used to test the dOTF
method by describing our high-dynamic range (HDR) imaging approach
(Section~\ref{sec:noisecomp}) followed by a description of the optical
setup (Section~\ref{sec:hardware}).


\subsection{Defocused high dynamic range imaging}
\label{sec:noisecomp}

In principle, the dOTF method could be implemented by taking two
consecutive images that have a very localized diversity near the pupil
border. In practice, we found that obtaining a high SNR is the
biggest challenge. This is not surprising when considering
Eq.~\eqref{eq:pd1} where two almost identical images are
subtracted. Furthermore, those images are the focal plane images of a
point-source, which means that -- in the absense of large wavefront
aberrations -- most of the light is localized in the core of a small
PSF.

A large defocus during the dOTF measurements spreads the light more
evenly over the detector pixels. This improves the SNR and
dramatically increases the accuracy of the higher spatial
frequencies of the complex amplitudes.

\begin{figure}[hbtp]  \center
  \includegraphics{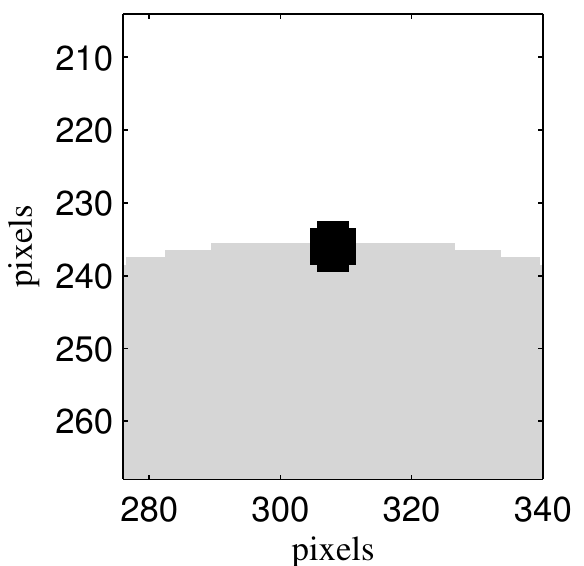}
  \caption{Illustration of the SLM mask we used to create the
    diversity for the dOTF method. Black circle shows the modified
    diversity pixels, gray pixels show the estimated pupil border.}
  \label{fg:divspot}
\end{figure}

In addition, we paid attention to other aspects that improve the
performance. Our image acquisition recipe for a single dOTF
measurement is:
\begin{itemize}
\item Create a defocus of $\sim$17--30~rad peak-to-valley. 
  Make sure the system is as stable as possible during the recording
  process. For instance, avoid using deformable mirrors with high
  voltages to reduce the effects of amplifier noise.

\item Adjust the laser power (or change neutral density filters) such
  that the camera can work with short exposure times (on the order of
  milliseconds). This reduces the effect of turbulence inside the
  optical setup and slowly drifting SLM characteristics.

\item Select optimal diversity. We used a disk having a width of
  2.5--4\% of the pupil diameter (corresponding to 5--8 SLM pixels).
  Place the center of the diversity at the detected pupil border (half
  of the disk inside the aperture, as illustrated in
  Fig.~\ref{fg:divspot}). A small disk makes the SNR low, a wide disk
  blurs and biases the recorded complex amplitudes. We used the SLM to
  introduce the maximum possible complex amplitude change within the
  diversity disk: 1.6~rad phase difference and 25\% drop in
  transmittance -- a value also found in \cite{codona2013}.

\item Record a set of images with varying exposure times such that
  long-exposure images saturate. Create compiled images (as discussed
  later) to further increase the camera's dynamic range and reduce
  its read-out noise. This is also called high dynamic range imaging
  (HDR) \cite{mann94}. We used 5--6 different exposure times
  (0.24--29~ms) for one compiled image.

\item Create a set of image compilations and average to
  further reduce the noise. We did this pair-wise: first one reference
  image compilation, then one diversity image compilation, then again
  one reference and so on. This avoids issues with internal turbulence
  (caused by air convection) and slow drifts of the SLM. We typically used
  40 such pairs for a single dOTF determination.
\end{itemize}
Our total recording time for one dOTF reconstruction was 2--4~min.

To create the compilation with increased dynamic range, we take a set
of $N$ images ($I_1$, $I_2$, $\ldots$, $I_N$), ordered according to 
increasing exposure time. We start by setting
$I=I_1$. For the remaining images ($k=2,\ldots, N$), we iteratively
update the compilation according to:
\begin{enumerate}
\item Calculate a scaling factor
  \begin{displaymath}
    s_k = \frac{\sum_{\Omega_1} I}{\sum_{\Omega_1} I_k},
  \end{displaymath}
  where $\Omega_1$ are the unsaturated pixels with adequate signal in
  both the current compilation ($I$) and $I_k$.

\item Update the compilation: $I\left\{\Omega_2\right\}$ = $s_k
  I_k\left\{\Omega_2\right\}$, where $\Omega_2$ is the set of pixels
  that are not saturated in $I_k$.
\end{enumerate}
The linear range and the saturation of the camera are determined by
inspecting the response when gradually increasing the exposure
time. To be on the safe side, we assume the linear range to be
15--3500 analog-to-digital units (ADUs) while the maximum intensity of
the camera is 4096 ADUs.

\subsection{Optical setup}
\label{sec:hardware}

\begin{figure}[hbtp]  \center
  \includegraphics[width=8.6cm]{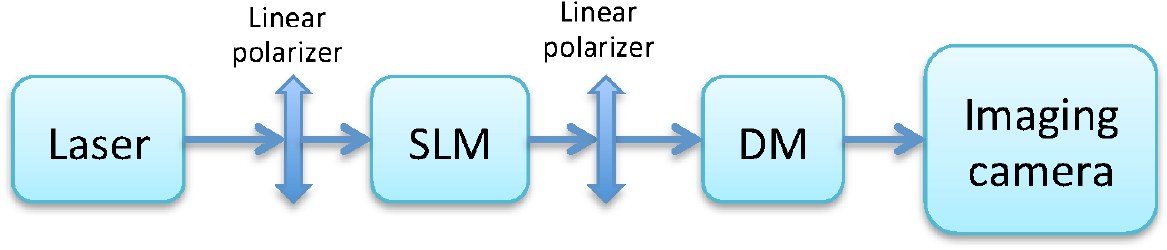}
  \caption{Schematic view of our optical setup.}
  \label{fg:schema}
\end{figure}

A schematic view of the optical setup is shown in
Fig.~\ref{fg:schema}. The beam is directed through a linear polarizer,
the spatial light modulator and another linear polarizer. Then, it is
reflected by a deformable mirror (DM) and focused onto the camera.

The rotation of the polarizers with respect to the SLM is chosen
such that the SLM produces maximum phase shifts and minimum
transmittance changes (as explained for instance in
\cite{remenyi2003}).

The light source is a laser diode having a wavelength of 656~nm.
The laser is coupled to a fiber, and the light is roughly 50\%
polarized when arriving at the first collimating lens. We found that
the quality of the laser was not a limiting factor,
and we made no attempts to study its stability or install a spatial
filter in front of the laser window.

Using standard 1~inch doublet lenses, the beam is collimated, passes a
diaphragm $\sim$5~mm wide and is re-collimated to pass
the SLM and DM with a diameter of $\sim$1~cm.

The SLM is a transmissive device having $800\times 600$
pixels. The birefringence of each SLM pixel can be controlled by a VGA
signal having a range of 8 bits per pixel (grayscale values between
0--255). For the experiments reported in this paper, we adjusted the
size of the diaphragm such that the beam passes an area of 334~pixels
wide.

The SLM causes a strong diffraction effect since the pixels and wiring
between them act as a grating. Several sub-beams emerge, but only the
brightest hits the camera. We observe no adverse effects from the
other beams.

In addition to the SLM, we use the DM to control the low-order modes of
the wavefront. During the dOTF measurements, the DM is not needed, and
we turn it off. However, when testing algorithms to achieve a flat
wavefront (see Section~\ref{sec:ff}), we use the DM to
overcome the stroke limitations of the SLM.

The whole setup was built, for convenience, on a $30\times
45$~cm breadboard with two folding mirrors (before the 1st polarizer
and before the imaging camera).  The detailed parameters are listed in
Table~\ref{tb:hardware}.

\begin{table}[hbtp] \begin{center}
  \caption{Details of used hardware}
  \label{tb:hardware}
  \begin{tabular}{ll} 
  \hline
  \hline
  \multicolumn{2}{c}{Imaging camera} \\
  Model                   & Basler piA640-210gm \\
  Resolution              & $648 \times 488$~pixels \\
  Dynamic range           & 12 bits \\
  Pixel size              & $7.4\times 7.4$~$\mu$m \\
  Readout noise           & 14 electrons \\
  Sensor type             & CCD \\
  \hline
  \multicolumn{2}{c}{Spatial light modulator} \\
  Model                   & Holoeye LC2002 \\
  Resolution              & $800\times 600$ pixels \\
  Fill-factor             & 55\% \\
  Dynamic range           & 8 bits \\
  Pixel pitch             & 32~$\mu$m \\
  \hline
  \multicolumn{2}{c}{Deformable mirror} \\
  Manufacturer            & Flexible Optical B.V. \\
  Type                    & Micromachined membrane DM \\
  Diameter                & 15 mm \\
  Number of channels      & 37 \\
  Controlled modes        & 36 \\
  \hline
  \multicolumn{2}{c}{Light source} \\
  Model                   & Qphotonics QFLD-660-2S \\
  Type                    & Laser diode, fiber coupled\\
  Central wavelength      & 656~nm\\
  Spectral bandwidth      & 0.7~nm \\
  \hline
  \end{tabular}
\end{center} \end{table}

\section{Results}
\label{sec:results}

This section describes the results of our laboratory experiments.
Section~\ref{sec:dotf0} illustrates the properties of the basic dOTF
method. Sections~\ref{sec:phresp} and \ref{sec:rotreg} show the
details of how we determined the SLM response and registration
with respect to the pupil. Finally, Section~\ref{sec:ff} describes the
experimental verification of a focal-plane wavefront sensing
algorithm.



\subsection{Basic features of dOTF arrays}
\label{sec:dotf0}

First, we show a basic example of the dOTF method when the DM is turned
off and all the SLM pixels are set to zero.  We also optimized the
camera focus and neutral density filters such that short exposure
times (0.24~ms) were possible with a large defocus ($\sim$30~rad
peak-to-valley) without saturating the camera.

We followed the procedure described in Section~\ref{sec:noisecomp}
with 5 different exposure times (0.24--4~ms). In this way, the dynamic
range of the images was $1 \cdot 10^7$~to~$5\cdot 10^7$ (ratio of
smallest and largest non-zero values). For convenience, we cropped
the central region of $320\times 320$ pixels from these HDR images; 
that region contains enough information for the dOTF reconstruction.
Then, we zero-pad the cropped parts to obtain arrays of $640\times 640$ 
pixels to avoid FFT wrap-around effects.

\begin{figure}[hbtp]
  \includegraphics{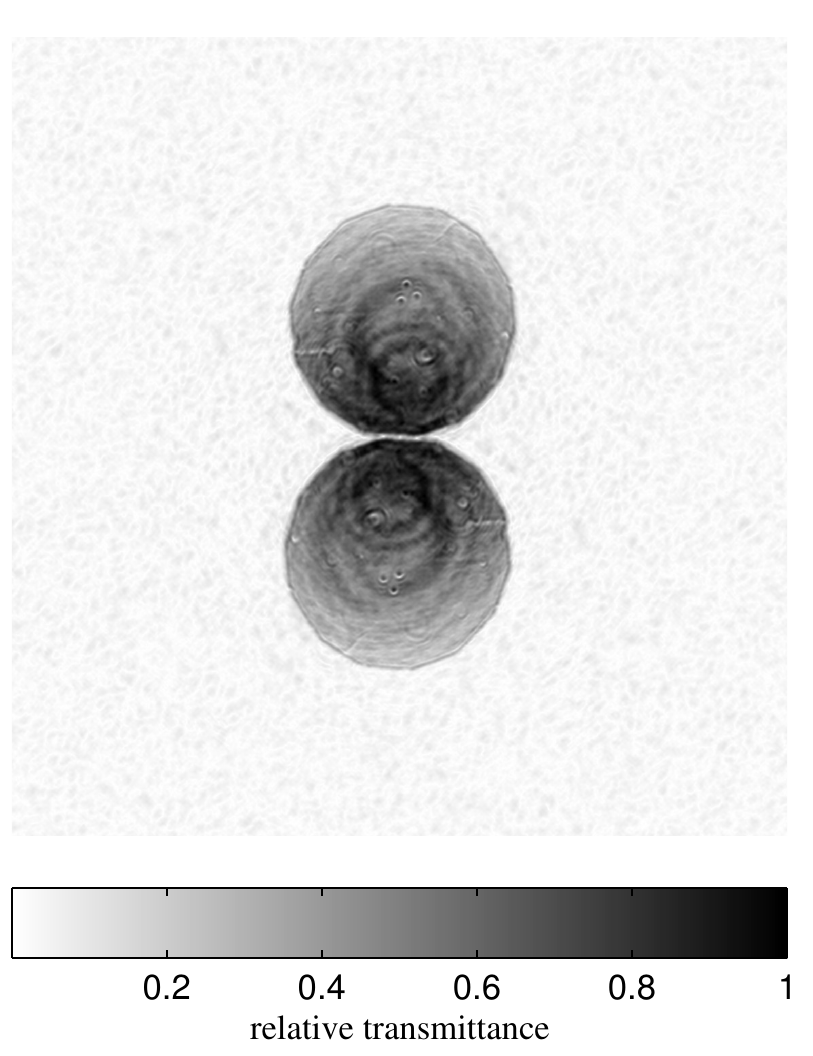}
  \includegraphics{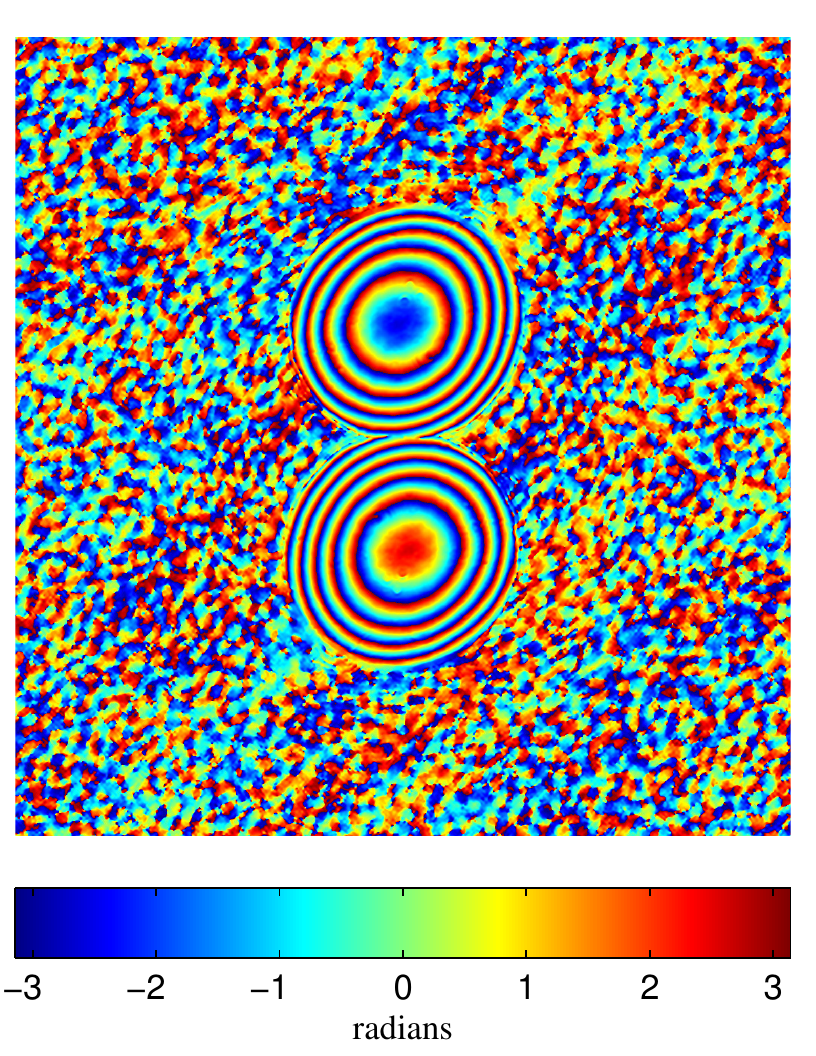}
  \caption{One dOTF used for the SLM calibration. The arrays
    are $640\times 640$ pixels wide. Upper: modulus. Lower: phase.}
 \label{fg:dotf0}
\end{figure}

The dOTF arrays were then computed as discussed in
Section~\ref{sec:basics}. We applied no windowing function to avoid
blurring the fine features, in particular the pupil borders. An
example of the resulting arrays is shown in Fig.~\ref{fg:dotf0}.

The dOTF modulus very clearly shows the same pattern as seen in the
simulated example in Section~\ref{sec:basics}.  Due to measurement
errors, additional noise is seen. Particularly outside of the circular
pupil, the noise can easily be characterized: it consists of small
clots, 4--7 pixels wide. The noise rms is $\sim$3.5\% of the maximum
intensity, and its peaks are always lower than $\sim$10\% of the
maximum. Ideally, the noise should be almost white since the PSF
measurements are limited by the read-out noise. However, the
zero-padding, the SLM drift and the internal turbulence cause more
complicated errors.

The separation of the two pupils is very clear, which indicates that
the applied diversity was close to a delta function. The very
localized diversity makes it possible to see fine-scale structures:
the shape of the used diaphragm, location and impacts of several dust
particles and larger, ring-like structures, whose origin is not clear
at this time.

In comparison to the simulated dOTF with a perfect pupil, as discussed
in section \ref{sec:esterr}, the border effect (a bright ring with
lower intensity inside the pupil) is an expected diffraction
effect. The measurement bias is also very similar to what is seen in
the simulation: the pupil intensity is highest close to the diversity
location, decreasing radially from that point.

The width of the pupil, compared to the dOTF array, is easy to
determine from the data in Fig.~\ref{fg:dotf0}. The intensity
increases from the noise level to the mean pupil intensity ($\sim$0.5
of the maximum) in $\sim$4~pixels. The pupils shown in
Fig.~\ref{fg:dotf0} have a width of 186$\pm$1~pixel, and therefore the
sampling, as defined by Eq.~\eqref{eq:sample}, can be determined with
a resolution of $\sim$1\% to be 3.43.

Fig.~\ref{fg:dotf0} also shows the phase in the pupil plane. Since we
applied a large defocus, the most dominant feature is the concentric
rings, caused by phase wrapping. We found that the noise levels were
sufficiently small such that the original wavefronts were extremely
easy to reconstruct by unwrapping the phase. To unwrap, we used the
standard quality-guided algorithm, discussed for instance in
\cite{GP1998}.

\subsection{Determining SLM phase and amplitude response}
\label{sec:phresp}

To measure the SLM response to a voltage change, we kept half of the
SLM pixels at zero, and the control signal of the other pixels was set
to a constant value. The dOTF method was used to determine the change
of the pupil plane complex amplitudes in the region where the SLM was
nonzero.

We noticed that the drift of the SLM created a significant change at
lower spatial frequencies, especially tip/tilt, at time scales of
$\sim$5--10 minutes. Therefore, we had to record a new reference
measurement (with all SLM pixels set to zero) separately for all the
SLM modifications we wanted to determine.

\begin{figure}[hbtp] \center
  \includegraphics{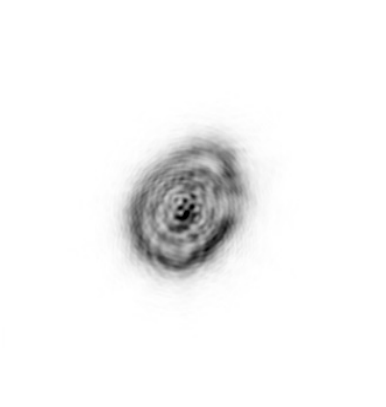}
  \includegraphics{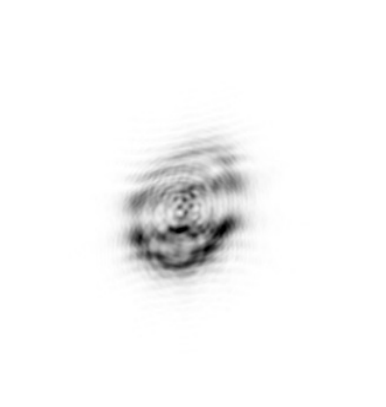}

  \includegraphics{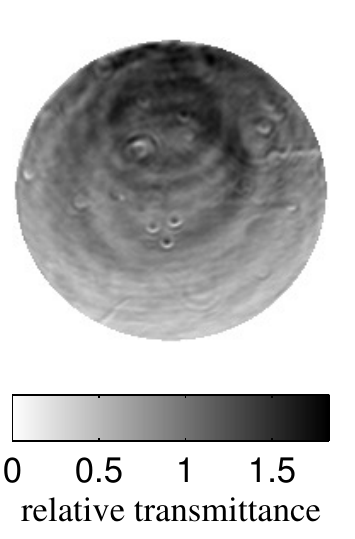}
  \includegraphics{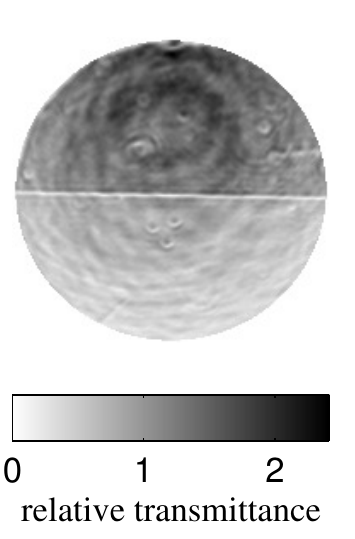}

  \includegraphics{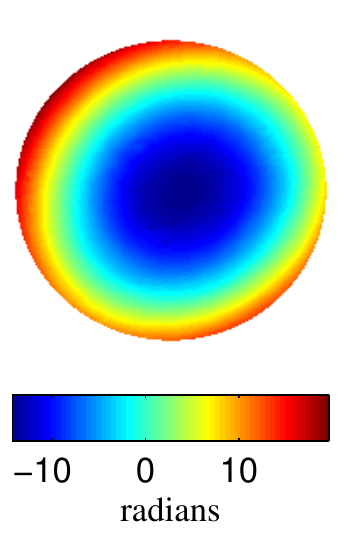}
  \includegraphics{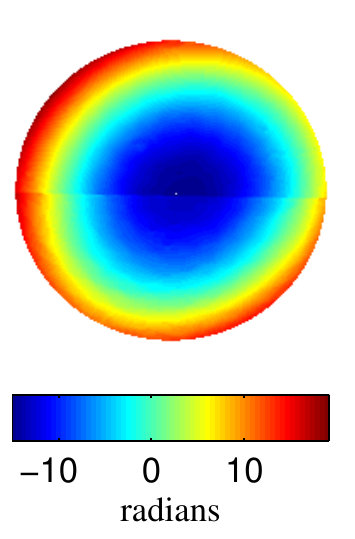}
  \caption{Examples of SLM calibration. Upper row: recorded
    images. Middle row: dOTF moduli. Lower row: unwrapped dOTF
    phase. Left column: reference (all SLM pixels at zero). Right
    column: half of SLM pixels set to maximum control voltage.}
 \label{fg:slmrespo1}
\end{figure}

Figure~\ref{fg:slmrespo1} shows the measured data and dOTF
reconstructions for two cases: the reference and a case where the SLM
control voltage is maximum in the lower part of the pupil.  The area
where the SLM pixels have been modified is clearly visible in the dOTF
arrays. The dOTF modulus has a clear edge with a reduced intensity
along the modification, exactly as in the simulations in
Section~\ref{sec:esterr}.

\begin{figure}[hbtp]
  \includegraphics{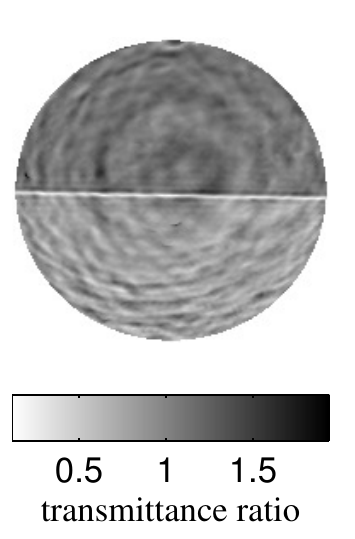}
  \includegraphics{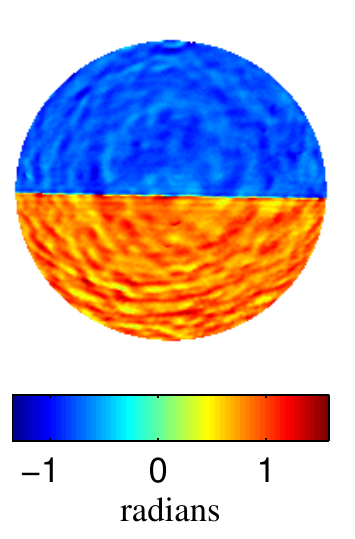}
 \caption{Examples of SLM calibration. Left: ratio of dOTF moduli in
   Fig.~\ref{fg:slmrespo1}. Right: difference of dOTF phase in
   Fig.~\ref{fg:slmrespo1}.}
 \label{fg:slmrespo2}
\end{figure}

Fig.~\ref{fg:slmrespo2} shows the arrays we used to determine the SLM
change. As in Section~\ref{sec:esterr}, we subtracted the unwrapped
dOTF phases to calculate the SLM phase change, and we took the ratio
of the dOTF moduli to calculate the SLM transmittance change.  To
reduce the impact of noise, we applied median filtering (window size
5) to the unwrapped phase and the dOTF moduli. We also masked out
2\% of the pupil at the edges to avoid dealing with the border
effects. The masking was done only to compare the average values of
the different half-pupils -- it has no impact on the other aspects of
the dOTF method.


Figure~\ref{fg:slmrespo2} shows an obvious change in the lower half of
the pupil. Compared to the simulated case in Fig.~\ref{fg:errsimu},
additional noise-like structure is present: the rms error of the upper
and lower semi-pupils are 0.09~rad and 0.16~rad, respectively, while
the average difference between the semi-pupils is 1.6~rad. For the
transmittance change, the values are 0.09 and 0.16, and the difference
between the semi-pupils is 0.25.



It is a reasonable approximation that all the SLM pixels have an
identical response and that the pixels are evenly distributed across
the screen. Thus, the additional structure is caused by measurement
errors. The drift of the SLM, together with a large defocus, can
create the ring-like structures, and internal turbulence along the
optical path is probably the reason for the speckle-like structure of
the errors. Also Fresnel propagation effects from the SLM to the last
pupil plane could have an impact, which would appear as additional
border effects around the edge of the SLM modification.

Nevertheless, a good estimate of the average change can be obtained by
taking an average over the whole area where a pupil modification is
observed.

\begin{figure}[hbtp]
\begin{tikzpicture}
\begin{pgfonlayer}{foreground}
     \path  (-1.8,2.1) node (c) {{Transmittance}}
        (-1.7,-0.8) node (b) {{Phase difference}};
\end{pgfonlayer}
\begin{pgfonlayer}{background}
 \path      (0,0) node (o) {
   \includegraphics[width=8.5cm]{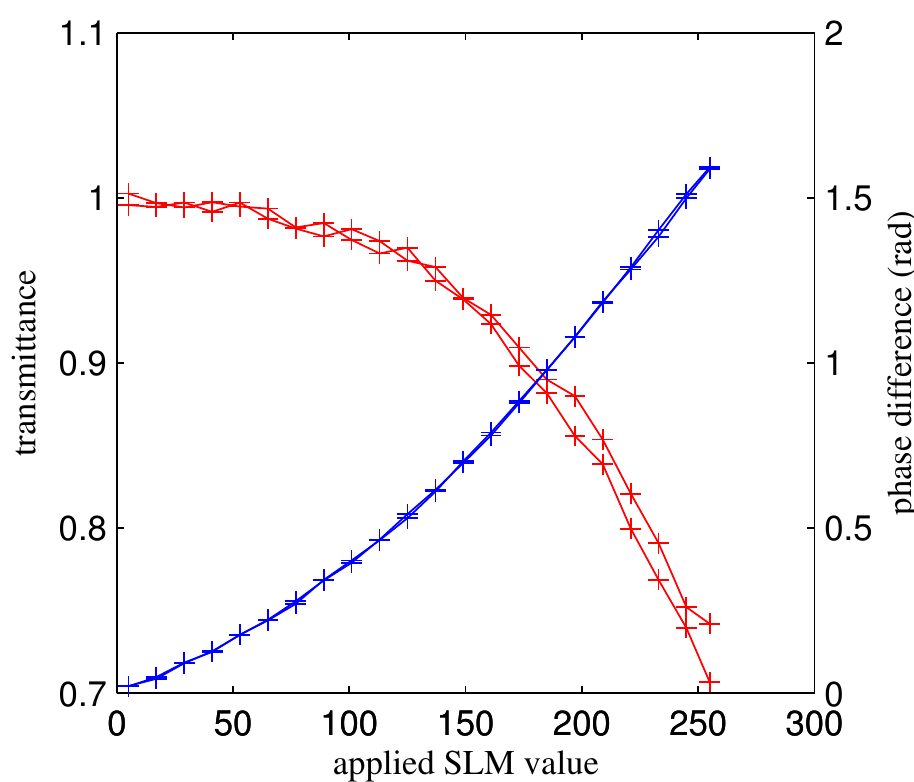}};
\end{pgfonlayer}
\end{tikzpicture}
\caption{SLM calibration results.  Phase and transmittance as a
  function of applied SLM control signal value. Two independent
  measurement series are shown. Note that the phase plots overlap.}
\label{fg:slmrespopl}
\end{figure}

We repeated the measurements shown in Figs.~\ref{fg:slmrespo1} and
\ref{fg:slmrespo2} for several SLM control levels. Two independent
series were recorded with different dOTF diversity locations. The
resulting transmittance and phase difference are shown in
Fig.~\ref{fg:slmrespopl}. The independent measurement series are in
excellent agreement: the difference of the two measurements is
typically $\sim$1\% of the average measurement. The transmittance is
more difficult to measure with higher accuracy, when a large SLM
control signal difference is applied; the difference grows to
$\sim$3--5\% at signals larger than 200. For the phase, the difference
between two independent measurements is always less than 0.005~rad
(excluding the control signal 233, where it is 0.01~rad).  This
measured SLM response is in excellent agreement with the results
reported in \cite{remenyi2003}.

The maximum observed stroke was 1.6~rad, but it comes at the cost of
$\sim$25\% loss in transmittance. If the maximum transmittance loss is
restricted to 15\%, the available stroke is $\sim$1.2~rad.


\subsection{Determining SLM registration}
\label{sec:rotreg}

To determine the SLM registration with respect to the physical pupil,
we first re-optimized the camera focus and the neutral density filters
such that the same setting could be used later with our focal-plane
wavefront-sensing algorithm. With these changes, the shortest possible
exposure time (80~$\mu$s) results in images at the optimal focus
having a peak intensity $\sim$50\% of the saturation level.

We applied the dOTF method with 6 different exposure times (2--29~ms)
with a defocus of $\sim$17~rad peak-to-valley, which was created by
the DM.  The resulting dynamic range of the recorded images was
similar to what we reported in the previous section, although with
larger errors caused by the longer integration times and smaller
defocus giving less light on the peripheral camera pixels. However, we
found the impact of these additional errors to be negligible for our
purposes. Yet, a significant change, caused by the refocusing, was
that the PSF sampling changed by $\sim$5\% to a value of $w=3.26$ due
to the lack of telecentricity of the beam reaching the camera.

In the same way as in the previous section, we introduced a change in
phase with selected SLM pixels. We used the maximum control signal
on stripe 66 pixels wide. We made 16 dOTF recordings
using both vertical and horizontal stripes, which covered the pupil at
equal intervals.

\begin{figure}[hbtp]
  \includegraphics{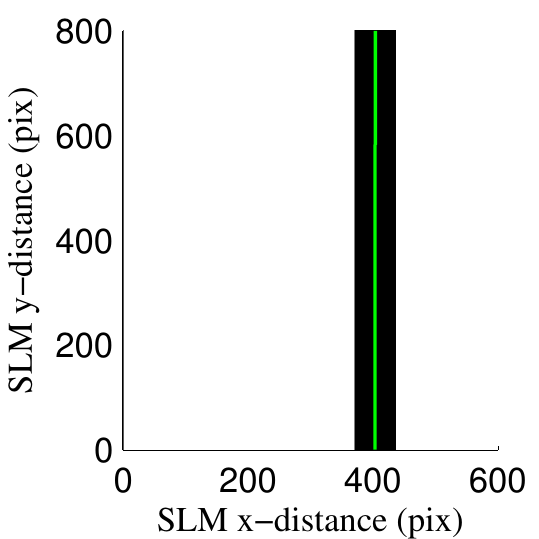}
  \includegraphics{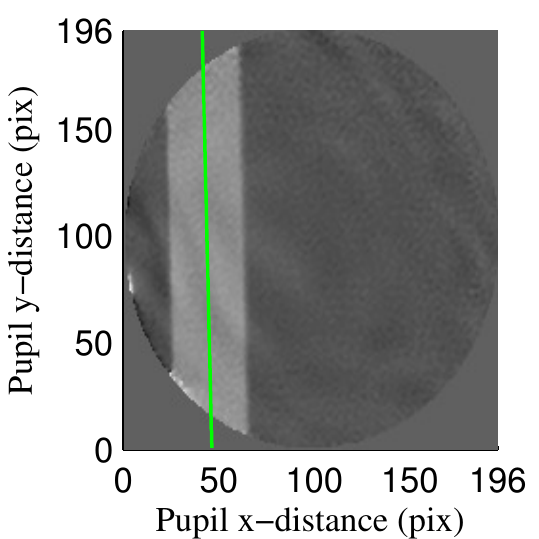}
  \includegraphics{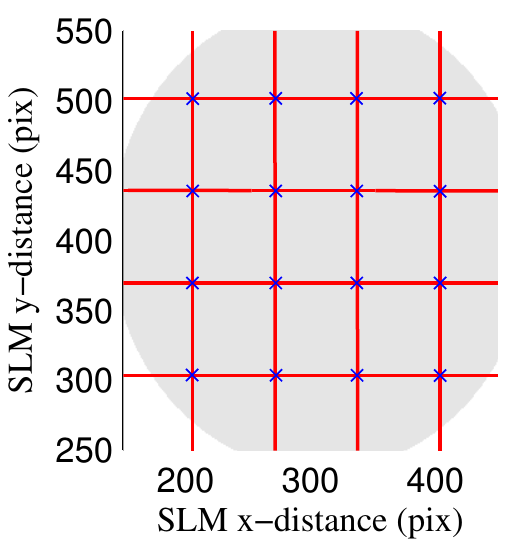}
  \caption{An example of the SLM registration determination. Top: SLM
    pixels that were modified and a linear fit along its
    center. Middle: pupil phase recorded by the dOTF method and a linear
    fit. Bottom: red lines show all the SLM pixel
    modifications. Crosses show the intersections of dOTF measured
    lines mapped onto the SLM coordinates using an optimal affine
    transform. Gray area shows estimated pupil location.}
 \label{fg:slmregi}
\end{figure}

Then, we determined the location of the modified stripe in the pupil
plane.  We found it easiest to do this by using a gradient
detection. The x- and y-gradients of the dOTF phase were computed by
subtracting two arrays, both shifted one pixel in opposite
directions. Then, we took a squared sum of the x- and y-gradients and
filtered out 98\% of the lowest values, thus keeping only the
pixels describing the borders of the stripe. The pixels were used to
compute two linear fits of the two edges of the stripe. Finally,
a mean of these two lines was calculated, and it accurately represents
the middle of the stripe.  An example of this is shown in
Fig.~\ref{fg:slmregi}.

The fitting produces two sets of intersections -- two grids of 16
points.  One of the sets is in the SLM pixel coordinates, the other in
the physical pupil coordinates (whose sampling is determined by the
size of the dOTF array).

Then, we calculate the optimal affine transform ($\affi$) minimizing
the error
\begin{equation}
  e = \sum_{i=1}^{16} \left\| \affi \left[ \begin{array}{c}
      x_\text{meas}(i) \\
      y_\text{meas}(i) \\
      1
    \end{array} \right] 
  -
  \left[ \begin{array}{c}
      x_\text{SLM}(i) \\
      y_\text{SLM}(i) \\
    \end{array} \right]
  \right\|^2,
\end{equation}
where $(x_\text{meas}(i), y_\text{meas}(i))$ are the coordinates
determined from the dOTF reconstructions, $(x_\text{SLM}(i),
y_\text{SLM}(i))$ are the SLM pixels corresponding to the
intersections determined in the pupil coordinates, $i$ refers to the
intersections of the fitted lines, and $\affi$ is the transform
expressed by a $2\times 3$ matrix. 

We used dOTF arrays having a size of $640\times 640$ pixels, and
therefore the pupil coordinates $(x_\text{meas}(i), y_\text{meas}(i))$
are within an array of $196\times 196$~pixels. That area maps to the
SLM pixels having a size of $334\times 334$ pixels and located within
the device with $800\times 600$ pixels. Thus, the mapping performs the
interpolation that defines how many degrees of freedom we actually use
to control the SLM.

We found that the intersections we determined in pupil coordinates can
be mapped to SLM pixel coordinates with sub-pixel accuracy: the
maximum error was 0.6~pixels, and the rms error was 0.4~pixels. The optimal
transform was
\begin{displaymath}
\affi =\left[ \begin{array}{rrr}
    1.687  & -0.029 & 235.9 \\
    -0.044 & -1.704 & 483.1 \\
\end{array}\right],
\end{displaymath}
which shows that a pure rotation, translation and scaling would be
sub-optimal to describe the SLM position with respect to the pupil
plane: the difference to the optimal translation would be about
1--2~SLM pixels at the peripheral pupil points.  This is probably caused
by a small tilt of the SLM or camera, compared to the optical axis.
The values show that the pixel grids of the SLM and camera are
rotated approximately 1--1.5$^\circ$ with respect to each other, and
this is also visible in Figs.~\ref{fg:slmrespo1}, \ref{fg:slmrespo2}
and \ref{fg:slmregi}.


After having determined $\affi$ and the SLM response (as in
Fig.~\ref{fg:slmrespopl}), it is possible to employ standard image
processing techniques to calculate the SLM control signal to create a
desired phase change in the physical pupil.


%


\subsection{Performance with a focal plane sensing algorithm}
\label{sec:ff}


Finally, we tested how a phase-diversity based focal-plane wavefront
sensing algorithm worked with a wavefront corrector calibrated as
described before; PSF sampling determined as discussed in
Section~\ref{sec:sampling}, SLM phase response and alignment
determined as shown in Sections~\ref{sec:phresp} and \ref{sec:rotreg},
respectively. We use the Fast \& Furious algorithm due to its easy
implementation.  Our earlier work \cite{korkiakoski2012spie1} has
illustrated how it works with low-order modes correction, and here we
show the results with a wavefront corrector having a spatial
resolution of $196\times 196$ control elements.

\begin{figure}[hbtp]
  \includegraphics[width=8.6cm]{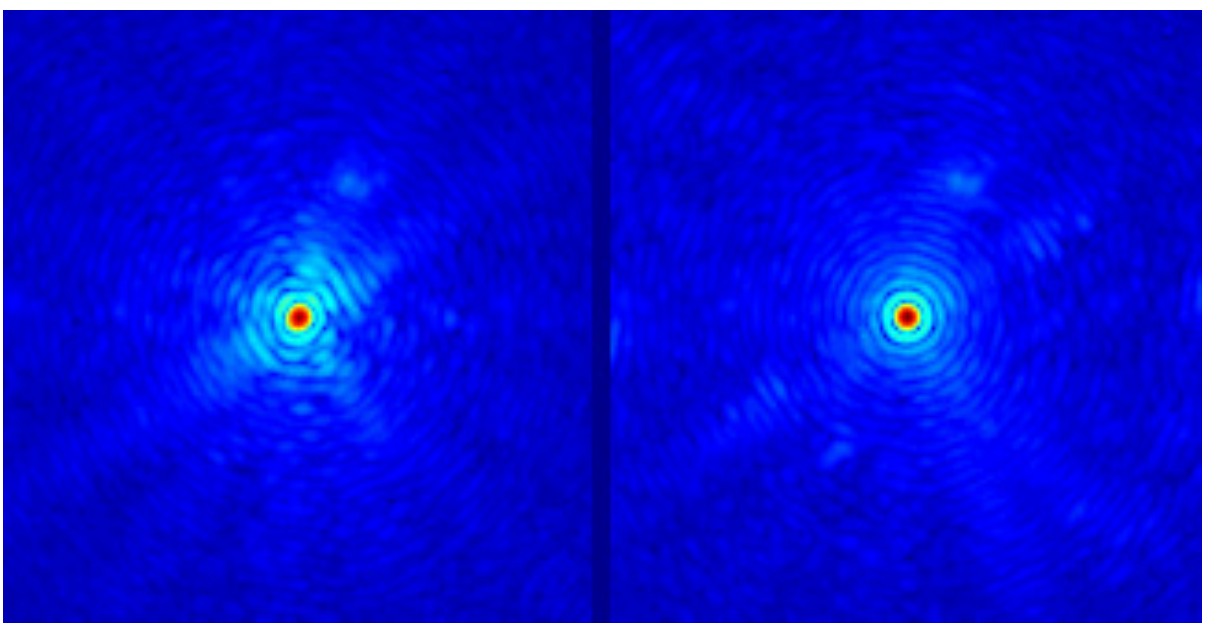}
  \caption{Illustration of the Fast \& Furious performance: PSF images
    raised to the 0.2 power. The images have a size of $160\times
    160$~pixels. Left: before SLM correction (Strehl ratio
    0.80). Right: after F\&F has converged (Strehl ratio 0.93).}
 \label{fg:ff}
\end{figure}

Fig.~\ref{fg:ff} illustrates the performance of F\&F. The left image
was obtained when the camera was well focused, and low-order modes
were corrected by the deformable mirror, but all the SLM pixels were
set to zero. The right image shows the situation after the F\&F
algorithm has converged. These two images -- and all the images used
by the algorithm -- are HDR compilations of 15 images with different
exposure times (0.08--64~ms). The improvement in the image quality is
obvious: the Strehl ratio (measured from the maximums of normalized
images) increased from 0.80 to 0.93.


The remaining errors are likely caused by a ghost (internal reflection
from SLM), saturation of the SLM (we limited the stroke to 1.2~rad
causing 16--20\% of the pixels to be saturated), and amplitude errors
(caused by both the SLM, imperfectly modeled pupil and Fresnel
propagation effects).  However, a more detailed analysis is outside
the scope of this paper, and we will address the issues in more detail
in an upcoming publication.

\section{Conclusions}
\label{sec:conclusions}

%
%

We have shown that an extremely straightforward method -- the
differential OTF -- can be used to calibrate a phase-diversity
based focal-plane wavefront sensing algorithm that corrects the
wavefront at a resolution of $\sim$$200\times 200$.

The dOTF method relies on localized diversities at the pupil border,
which is a unique feature compared to other focal-plane wavefront
sensing techniques. It requires no complicated hardware, physical
movement of optical components or demanding numerical computations,
and it therefore is an excellent option when an easily implementable
way to accurately determine the pupil-plane electric field with a
focal-plane camera is needed.

We have outlined how to theoretically predict the performance of
the method, and our simulations are able to explain the
experimental results -- apart from the instability issues related to our
hardware.

Based on theorical reasoning, we know that the ultimate resolution of
the dOTF method is limited by the size and shape of the used
diversity; it determines the convolution kernel that blurs the
original complex amplitudes.  In our experiments, the used diversity
gives a maximum resolution between $\sim$$50\times 50$ and
$\sim$$150\times 150$~pixels -- but the resolving power can be
different in vertical and horizontal directions.  A visual inspection
is in agreement with these values, although read-out noise and
instabilities reduce the practical resolution of the instantaneous
dOTF arrays.

To increase the calibration accuracy, we have used the fact that our
wavefront corrector, an SLM, forms a fixed, rectangular array of
evenly distributed pixels. We used the dOTF method to detect the sharp
borders caused by modified wavefront blocks, and using those borders,
we calculated the optimal affine transform projecting the SLM pixel
locations onto the physical pupil plane. Based on the match of that
mapping, we conclude that the locations of the SLM pixels are
determined with an accuracy of 0.3\% with respect to the pupil
diameter.

This paper concentrates on concepts necessary for future extreme
adaptive optics systems like the one necessary for the direct
exoplanet imager ELT-PCS \cite{kasper2013}.  However, the methods
discussed here are very versatile, and they can be put into use in
many systems -- in principle everywhere where measurements of
optical aberrations are needed.

We have used the SLM to generate the localized diversities. However,
they could as well be created by high-resolution deformable mirrors
such as in \cite{poyneer2011}. Also, additional simple mechanics at the
pupil edge could be used to introduce a small, localized obstruction as
a diversity.

The technique is best suited for applications where a monochromatic
light source is available. If the spectral bandwidth is increased, the
diffraction features further off-axis are blurred, which implies
that higher spatial frequency information in the dOTF
reconstruction is lost. However, lower spatial frequencies can still
be recovered to some extent \cite{codona2013}. High-contrast
speckle-nulling experiments have demonstrated good success using
10\% bandwidth \cite{giveon2007spie}, and probably a similar
bandwidth is feasible also with the dOTF method.

Cases that will benefit from the dOTF calibration are test benches
demonstrating techniques for ultra-high contrast imaging needed in
space-based exo-planet detection (e.g., HCIT \cite{lowman2004}),
experiments for ground-based extreme adaptive optics (ExAO testbed
\cite{severson2006}, HOT \cite{vernet06}, FFREE \cite{antichi2010})
and the path-finder XAO instruments (GPI \cite{macintosh06}, SPHERE
\cite{fusco06sphere}).

The main issue with the dOTF method is the measurement noise. We need
a recording time of 2--4~min for each dOTF measurement to reduce the
read-out noise and achieve the required high dynamic range. During
this recording process, the system behavior should be
stationary. However, we observed the drift of the SLM when trying to
increase the measurement time to more than 5~min; a real DM would
likely be better. Internal turbulence in the optical setup can also be
a serious problem.

Another point of concern is the resolution of the pupil plane
measurement. It is ultimately limited by the size of the localized
diversity, and we proposed to create it with the wavefront corrector
elements. Thus, it is necessary that the elements are evenly spaced
across the pupil and that they have identical response. What we can
measure is limited to the response of larger blocks of the correction
elements -- when a typical deformable mirror is used to introduce the
diversity, the coupling of the actuators is an additional
challenge.

Our future work will concentrate on demonstrating how the dOTF method
can be used in a wider range of optical experiments. These include
additional optics such as apodizers and coronagraphs. In addition, we
will include a more detailed analysis of the light propagation effects
from the wavefront corrector to the last pupil plane.

\section*{Acknowledgments}
We thank Gerard van Harten and Tim van Werkhoven for assistance with
the spatial light modulator and optical experiments.
\vspace{0.2cm}

\bibliographystyle{osajnl}

\end{document}